\documentstyle[12pt]{article}

\newcommand{\be}{\begin{equation}}
\newcommand{\ee}{\end{equation}}
\newcommand{\dlt}{\delta}
\newcommand{\Dlt}{\Delta}
\newcommand{\ra}{\rightarrow}
\newcommand{\vp}{\varphi}
\newcommand{\bt}{\beta}
\newcommand{\al}{\alpha}
\newcommand{\prt}{\partial}
\newcommand{\Om}{\Omega}
\newcommand{\om}{\omega}

\newcommand{\lbd}{\lambda}
\newcommand{\gm}{\gamma}
\newcommand{\Gm}{\Gamma}

\newcommand{\ep}{\varepsilon}
\newcommand{\bA}{{\bf A}}
\newcommand{\br}{{\bf r}}
\newcommand{\bE}{{\bf E}}
\newcommand{\bH}{{\bf H}}
\newcommand{\bJ}{{\bf J}}
\newcommand{\bj}{{\bf j}}
\newcommand{\bP}{{\bf P}}
\newcommand{\bS}{{\bf S}}
\newcommand{\bd}{{\bf d}}
\newcommand{\bk}{{\bf k}}

\begin{document}

\begin{center}
{\Large{\bf Turbulent Filamentation in Lasers with High Fresnel
Numbers} \\ [5mm]

V.I. Yukalov} \\ [3mm]

{\it
$^1$Bogolubov Laboratory of Theoretical Physics, \\
Joint Institute for Nuclear Research, Dubna 141980, Russia \\
and \\
Institut f\"ur Theoretische Physik, \\
Freie Universit\"at Berlin, Arnimallee 14, D-14195 Berlin, Germany
}

\end{center}

\vskip 2cm

{\bf Running title}: Turbulent filamentation in lasers

\vskip 2cm

\begin{abstract}

The theory of turbulent photon filamentation in lasers with high Fresnel
numbers is presented. A survey of experimental observations of turbulent
filamentation is given. Theoretical description is based on the method of
eliminating field variables, which yields the pseudospin laser equations.
These are treated by the scale separation approach, including the
randomization of local fields and the method of stochastic averaging.
The initial, as well as the transient and final stages of radiation
dynamics are carefully analysed. The characteristics of photon filaments
are obtained by involving the probabilistic approach to pattern selection.

\end{abstract}

\newpage

\section{Introduction}

In nonlinear media interacting with electromagnetic fields there appear
different spatiotemporal structures that are analogous to the structures
arising in many other complex nonequilibrium systems [1--5]. The most
known among such electromagnetic structures are the optical filaments
which can be formed in passive nonlinear matter [3,4,6] and in active
laser media [1,4,5]. Examples of filaments occurring in other
nonequilibrium systems can be found in the reviews [2,5]. In addition,
it is possible to mention the generation of filaments in hadron plasma
[7] and in quark-gluon plasma formed under heavy nucleus collisions
[8,9].

The present survey will be focused on the optical filaments developing
in lasers. The behaviour and characteristics of these filaments essentially
depend on the value of the Fresnel number $F\equiv R^2/\lbd L$, in which
$R$ and $L$ are the internal radius, aperture radius, and effective length,
respectively, of a cylindrical laser, and $\lbd$ is the optical wavelength.
There are two types of optical filaments, regular and turbulent, corresponding
to either low or high Fresnel numbers. The difference between these two
filament types, their main properties, and the related experiments are
discussed in Sec. 2.

Our concern here is the turbulent filaments arising in lasers with high
Fresnel numbers. The transition from a regular filamentary structure to
the turbulent filament behaviour is somewhat analogous to a crossover
phase transition in statistical systems [9--11], or a more close analogy
is the transition from the laminar to turbulent motion in liquids [1,2,12].
The description of the turbulent motion is notorious to be rather difficult
and is usually done by invoking a kind of averaging procedure. Optical
turbulent structures also require the usage of averaging techniques.

First, the turbulent filamentation in laser media was described on the
basis of stationary models [13--16] invoking the notion of an effective
time-averaged energy. A more elaborated approach, based on realistic
evolution equations, was developed in Refs. [5,17--20]. It turned out
that for this purpose it is convenient to work not with the standard
Maxwell-Bloch equations but with the evolution equations for the
pseudospin operators, resulting after the elimination of the field
variables. Since this type of equations is less known for collective
electromagnetic phenomena, these equations are derived in Sec. 3, and
their specification for treating the turbulent filamentary structure
is given in Sec. 4.

Turbulent filamentation in lasers is a self-organized process due to the
photon exchange through the common radiation field. Coherent radiation
in filaments develops in a self-organized way, even when there has been
no coherence at the initial time. The triggering mechanism for the
origination of the radiation coherence is the existence of transition
dipolar waves caused by the dipolar part of effective atomic interactions.
This is emphasized in Sec. 5.

The overall dynamics of radiating filaments is analysed in Sec. 6,
starting from the quantum stage, through the transient flashing regime,
to the stationary state. Generally, the filaments of different radii can
emerge, whose classification is given by means of the probabilistic
approach to pattern selection [18,21,22]. This allows us to define the
typical filament radius and the number of filaments, which is done in
Sec. 7.

The present work does not merely summarize and review the previous
theoretical results on the turbulent filamentation in laser media but
also contains several improvements of the theory, making the latter
better grounded both mathematically and physically.

\section{Experiments on Filamentation in Lasers}

First of all, let us recall that there are two types of optical
filaments, regular and turbulent, which is related to the value of
the laser Fresnel number. The latter plays for optical systems the
same role as the Reynolds number for moving fluids. When increasing
the Reynolds number, a laminar fluid transforms to turbulent. In the
similar manner, increasing the Fresnel number makes a regular
filamentary structure turbulent. Optical turbulence implies, by analogy
with the fluid turbulence, that the spatiotemporal dynamics is chaotic.
This means that the radiating filaments are randomly distributed in space
and are not correlated with each other.

Optical filaments are observed in the near-field cross-section of
lasers. The typical picture, when varying the Fresnel number is as
follows. At very small Fresnel numbers $F\ll 1$, there exists the
sole transverse central mode uniformly filling the laser medium.
When the Fresnel number is around $F\sim 1$, the laser cavity
can house several transverse modes seen as a regular arrangement
of bright spots in the transverse cross-section. Each mode
corresponds to a filament extended through the cylindrical volume.
This filamentary structure is regular in space, forming ordered
geometric arrays, such as polygons. The transverse structure is
imposed by the cavity geometry, being prescribed by the empty-cavity
Gauss-Laguerre modes. Such regular structures are well understood
theoretically, their description being based on the field expansion
over the modal Gauss-Laguerre functions related to the cylindrical
geometry [1]. For Fresnel numbers up to $F\approx 5$, the number
of bright filaments follows the $F^2$ law as $F$ increases. The
regular filamentary structures have been observed in several lasers,
such as CO$_2$ and Na$_2$ lasers [1]. Similar structures also appear
in many passive nonlinear media, e.g. in Kerr medium [3,4] and in
active nonlinear media, as the photorefractive Bi$_{12}$SiO$_2$
crystal pumped by a laser [3,4].

As soon as the Fresnel number reaches $F\approx 10$, there occurs a
qualitative change in the features of the filamentary structure: The
regular filaments become turbulent. This transition goes gradually,
as a crossover, with the intermittent behaviour in the region
$5<F<15$. The character of this change is again common for lasers
[23,24] and for active nonlinear media [3,4].

For Fresnel numbers $F>15$, the arising filamentary structures
are principally different from those existing at low Fresnel numbers.
The spatial structures now have no relation to the empty-cavity modes.
The modal expansion is no longer relevant and the boundary conditions
have no importance. The laser medium houses a large number of parallel
independent filaments exhibiting themselves as a set of bright spots
randomly distributed in the transverse cross-section. The number of
these random filaments is proportional to $F$, contrary to the case
of low Fresnel numbers, when the number of filaments is proportional
to $F^2$. The chaotic filaments, being randomly distributed is space,
are not correlated with each other. Such a spatio-temporal chaotic
behaviour is characteristic of hydrodynamic turbulence, because of
which the similar phenomenon in optics is commonly called the {\it
optical turbulence}. In contrast to the regime of low $F$, where the
regularity of spatial structures is prescribed by the cavity geometry
and boundary conditions imposing their symmetry constraints, the
turbulent optical filamentation is strictly self-organized, with its
organization emerging from intrinsic properties of the medium. Since
the optical turbulence is accompanied by the formation of bright
filaments with a high density of photons, this phenomenon can be
named [17] the {\it turbulent photon filamentation}. This phenomenon
is common for lasers as well as for photorefractive crystals [3--5].

The first observations of the turbulent filamentary structures in
lasers, to my knowledge, were accomplished in the series of experiments
[25--29] with the resonatorless superluminescent lasers on the vapours
of Ne, Tl, Pb, N$_2$, and N$_2^+$. In these experiments, the typical
characteristics were as follows: $\lbd\approx 5\times 10^{-5}$ cm,
$R\approx 0.1-0.3$ cm, $L\approx 20-50$ cm, and $F\approx 10-100$. The
number of filaments was $N_f\sim 10^2-10^3$, with the typical radius
$r_f\approx 0.01$ cm.

Then the filamentary structures in large-aperture optical devices
have been observed in several lasers, as reviewed in [23,24], and in
photorefractive crystals [3,4]. Numerical simulations have been
accomplished [30]. Experimental works mainly dealt with the CO$_2$
lasers [23,24,31,32], dye lasers [33], and semiconductor lasers [34,35].

The turbulent nature of filamentation occurring in high Fresnel number
lasers was carefully studied in a series of nice experiments [36--42]
with CO$_2$ lasers and dye lasers. Irregular temporal behaviour was
observed in local field measurements. It was found that the transverse
correlation length was rather short. Randomly distributed transverse
patterns generated in short times were observed, being shot-to-shot
nonreproducible. For intermediate Fresnel numbers $F\sim 10$,
instantaneous transverse structures were randomly distributed in space,
but after being temporally averaged, they displayed a kind of regularity
related to the geometrical boundary conditions. This type of combination
of irregular instantaneous patterns with the averaged or stationary
pattern, showing the remnant ordering, is understandable for the
intermediate regime in the crossover region $5<F<15$. Fully developed
optical turbulence is reached as the Fresnel number increases up to
$F\sim 100$.

The typical laser parameters are as follows [36--42]. The pulsed CO$_2$
laser, with the wavelength $\lbd=1.06\times 10^{-3}$ cm and frequency
$\om=1.78\times 10^{14}$ s$^{-1}$, emits the pulses of $\tau_p\approx
0.7\times 10^{-7}$ s or $10^{-6}$ s. The aperture radius $R\approx 1$
cm, laser length $L=100$ cm. The inversion and polarization decay rates
are $\gm_1=10^7$ s$^{-1}$ and $\gm_2=3\times 10^9$ s$^{-1}$. The CO$_2$
density is $\rho=2\times 10^{18}$ cm$^{-3}$. The Fresnel number is
$F\approx 10$. The characteristic filament radius is $r_f\approx 0.1$
cm.

The pulsed dye laser, with the wavelength $\lbd=0.6\times 10^{-4}$ cm
and frequency $\om=3.14\times 10^{15}$ s$^{-1}$, produces pulses of
$\tau_p\approx 0.5\times 10^{-6}$ s. The decay rates are $\gm_1=4\times
10^8$ s$^{-1}$ and $\gm_2=10^{12}$ s$^{-1}$. The cavity length is
$L\approx 20$ cm. By varying the aperture radius between $0.3$ cm and
$0.8$ cm, the Fresnel number can be changed by an order, between $F=15$
and $F=110$. The typical filaments radius is $r_f\approx 0.01$ cm.

In a broad-aperture pulsed dye laser, two different spatial scales
were noticed [33], one order of magnitude apart. The appearance of
the second, larger, scale could be due to a nonvanishing interaction
between filaments.

\section{Pseudospin Laser Equations}

Spatiotemporal laser dynamics is usually treated in the frame of the
Maxwell-Bloch equations [43]. One commonly employs these equations for
defining the points of stability of uniform solutions, which means the
appearance of nonuniform structures. The linear stability analysis of
the Maxwell-Bloch equations is the maximum one can do in the case of
well developed optical turbulence.

Another approach is based on the elimination of field variables
and dealing solely with the equations for spin operators [5,44]. This
approach possesses the following advantages: (i) It is microscopic,
which allows us to better understand the underlying physics of collective
effects. (ii) It is quantum, which allows for the description, when
coherence has not yet been developed. (iii) It provides us the possibility
not solely for finding instability points but for describing the whole
dynamics under the condition of strong turbulence. Since this approach
is not widely known, the related evolution equations are derived below.

Aiming at pursuing a microscopic picture, let us start with a realistic
Hamiltonian
\be
\label{1}
\hat H = \hat H_a + \hat H_f + \hat H_{af} + \hat H_{mf}
\ee
representing a system of resonant two-level atoms plus electrodynamic
field. The atomic Hamiltonian is
\be
\label{2}
\hat H_a = \sum_{i=1}^N \om_0 \left ( \frac{1}{2} + S_i^z \right ) \; ,
\ee
where $N$ is the number of atoms; $\om_0$, carrying transition frequency;
$S_i^z$, spin operator of an $i$-th atom. Wishing to be more rigorous in
terminology, we should call the operators $S_i^\al$, appearing here and
in what follows, the pseudospin operators, since they correspond not
to actual spins but to the population difference and dipole transition
operators. But for brevity, one often calls them just spin operators,
which should not bring confusion. The field Hamiltonian is
\be
\label{3}
\hat H_f = \frac{1}{8\pi} \; \int \left (\bE^2 + \bH^2\right ) \; d\br
\ee
with electric field $\bE$ and magnetic field $\bH=\nabla\times\bA$.
The vector potential $\bA$ is assumed to satisfy the Coulomb calibration
\be
\label{4}
\nabla\cdot\bA = 0 \; .
\ee
The atom-field interaction is described by the Hamiltonian
\be
\label{5}
\hat H_{af} = -\sum_{i=1}^N \left ( \frac{1}{c}\; \bJ_i\cdot\bA_i +
\bP_i\cdot\bE_{0i}\right ) \; ,
\ee
in which the short-hand notation is used for the vector potential
$\bA_i\equiv\bA(\br_i,t)$ and the external electric field
$\bE_{0i}\equiv\bE_0(\br_i,t)$, and where the transition current is
\be
\label{6}
\bJ_i = i\om_0\left (\bd S_i^+ - \bd^* S_i^-\right )
\ee
and the transition polarization is
\be
\label{7}
\bP_i = \bd S_i^+ + \bd^* S_i^- \; ,
\ee
with the atomic transition dipole $\bd$ and the ladder operators
$S_i^\pm\equiv S_i^x \pm iS_i^y$. Except resonant atoms, the laser
cavity can contain some additional filling matter [45] interacting with
the electromagnetic field through the Hamiltonian
\be
\label{8}
\hat H_{mf} = -\frac{1}{c}\; \int {\bf j}_{mat}(\br,t)\cdot
\bA(\br,t)\; d\br \; ,
\ee
where ${\bf j}_{mat}$ is the local density of current in the filling
matter. Here and in what follows, we set $\hbar\equiv 1$.

The field operators satisfy the equal-time commutation relations
$$
\left [ E^\al(\br,t),\; A^\bt(\br',t)\right ] =
4\pi ic\dlt_{\al\bt}(\br-\br') \; ,
$$
\be
\label{9}
\left [ E^\al(\br,t),\; H^\bt(\br',t)\right ] =  - 4\pi ic
\sum_\gm \ep_{\al\bt\gm}\; \frac{\prt}{\prt r^\gm}\; \dlt(\br-\br') \; ,
\ee
in which $\ep_{\al\bt\gm}$ is the unitary antisymmetric tensor [46] and
the so-called transverse $\dlt$-function is
$$
\dlt_{\al\bt}(\br) \equiv \int \left ( \dlt_{\al\bt} - \;
\frac{k^\al k^\bt}{k^2}\right ) \; e^{i\bk\cdot\br}\;
\frac{d\bk}{(2\pi)^3} =
$$
\be
\label{10}
\dlt_{\al\bt} \dlt(\br) + \frac{\prt^2}{\prt r^\al\prt r^\bt} \;
\int \frac{e^{i\bk\cdot\br}}{k^2} \; \frac{d\bk}{(2\pi)^3} =
\frac{2}{3}\; \dlt_{\al\bt}\; \dlt(\br) - \; \frac{1}{4\pi} \;
D_{\al\bt}(\br) \; ,
\ee
where in the dipolar tensor
\be
\label{11}
D_{\al\bt}(\br) \equiv \frac{\dlt_{\al\bt}-3n^\al n^\bt}{r^3}
\ee
one has ${\bf n}\equiv\br/r=\{ n^\al\}$ and $r\equiv|\br|$.

The Heisenberg equations of motion for the field operators yield
\be
\label{12}
\frac{1}{c}\; \frac{\prt \bE}{\prt t} =\nabla\times\bH - \;
\frac{4\pi}{c}\; \bj\; , \qquad
\frac{1}{c}\; \frac{\prt\bA}{\prt t} = - \bE \; ,
\ee
from where, with the Coulomb calibration (4), one has the equation for
the vector potential
\be
\label{13}
\left ( \nabla^2 - \; \frac{1}{c^2} \; \frac{\prt^2}{\prt t^2}
\right ) \bA =  - \; \frac{4\pi}{c}\; \bj \; ,
\ee
with the density of current
\be
\label{14}
j^\al(\br,t) = \sum_\bt \left [ \sum_{i=1}^N \dlt_{\al\bt}(\br -\br_i)
J_i^\bt(t) + \int \dlt_{\al\bt}(\br-\br') j_{mat}^\bt(\br',t)\; d\br'
\right ] \; .
\ee
The known solution to Eq. (13) is the sum
\be
\label{15}
\bA(\br,t) = \bA_{vac}(\br,t) + \frac{1}{c}\; \int \bj
\left ( \br',t -\; \frac{|\br-\br'|}{c}\right )\;
\frac{d\br'}{|\br-\br'|}
\ee
of
the vacuum vector potential and the retarded potential.

The Heisenberg equations of motion for the spin operators, satisfying
the commutation relations
$$
\left [ S_i^+,\; S_j^-\right ] = 2\dlt_{ij} S_i^z \; , \qquad
\left [ S_i^z,\; S_j^\pm \right ] = \pm\dlt_{ij} S_i^\pm \;
$$
lead the equations
$$
\frac{dS_i^-}{dt} = -i\om_0 S_i^- + 2 S_i^z \left ( k_0\bd\cdot\bA_i -
i\bd\cdot\bE_{0i}\right ) \; ,
$$
\be
\label{16}
\frac{dS_i^z}{dt} = - S_i^+ \left ( k_0\bd\cdot\bA_i -
i\bd\cdot\bE_{0i}\right )  - S_i^-\left ( k_0\bd^*\cdot\bA_i +
i\bd^*\cdot\bE_{0i}\right ) \; .
\ee
These are complimented by the retardation condition
\be
\label{17}
S_i^\al(t) = 0 \qquad (t<0) \; .
\ee

In order that the notion of resonant atoms would have sense,
one requires, as usual, that the atom-field interactions are small
compared to the atomic transition energy, so that $|\bd\cdot\bE|
\ll\om_0$. Because of this, the retardation can be taken into account
in the Born approximation
\be
\label{18}
S_j^- \left ( t -\; \frac{r}{c}\right ) =
S_j^-(t)\Theta(ct-r)\; e^{ik_0r} \; , \qquad
S_j^z \left ( t -\; \frac{r}{c}\right ) = S_j^z(t)\Theta(ct-r) \; ,
\ee
where $k_0\equiv\om_0/c$ and $\Theta(t)$ is the unit-step function.

The idea of eliminating the field operators is based on the usage of
the pseudospin equations (16), with the substituted there the vector
potential (15). In this way, one meets the terms corresponding to the
atomic self-action, which in the present approach can be treated as
follows. The vector potential generated by a single atom is
\be
\label{19}
\bA_s(\br,t) = \frac{1}{c}\; \int
\frac{\dlt_{\al\bt}(\br')}{|\br-\br'|} \;
J\left ( t -\; \frac{|\br-\br'|}{c}\right ) \; d\br' \; ,
\ee
with the current
$$
\bJ\left ( t - \; \frac{r}{c}\right ) = i\om_0\left [
\bd S^+(t)\; e^{-ik_0r} -\bd^*\; S^-(t)\; e^{ik_0r}\right ]\;
\Theta(ct-r) \; ,
$$
where $S^\al(t)\equiv S^\al(0,t)$. At small distance, such that
$k_0r\ll 1$, one may write $e^{ik_0r}\simeq 1+ik_0r$. Substituting the
transverse $\dlt$-function (10) into the vector potential (19), we keep
in mind that averaging the dipolar tensor (11) over spherical angles
gives
$$
\int D_{\al\bt}(\br) \; d\Omega(\br) = 0 \; .
$$
Then, for $k_0r\ll 1$, the vector potential (19) becomes
\be
\label{20}
\bA_s(\br,t) \simeq \frac{2}{3}\; k_0^2\left [ \bd S^+(t) +
\bd^* S^-(t)\right ] + i\; \frac{2k_0}{3r}\; \left [
\bd S^+(t) - \bd^* S^-(t)\right ] \; .
\ee
To avoid the divergence in the term $1/r$, let us average it between the
electron wavelength $\lbd_e=2\pi\hbar/mc$, with $m$ being the electron
mass, and the radiation wavelength $\lbd_0=2\pi/k_0$. Taking into account
that $\lbd_e\ll\lbd_0$, we have
$$
\frac{1}{\lbd_0-\lbd_e} \; \int_{\lbd_e}^{\lbd_0} \frac{dr}{r} =
\frac{k_0}{2\pi}\; \ln\left ( \frac{mc^2}{\hbar\om_0} \right ) \; .
$$
Then for the self-acting vector potential, we get
\be
\label{21}
\bA_s(0,t) = \frac{2}{3}\; k_0^2 \left [ \bd S^+(t) + \bd^* S^-(t)
\right ] + \frac{ik_0}{3\pi}\; \ln\left ( \frac{mc^2}{\hbar\om_0}\right )
\left [ \bd S^+(t) - \bd^* S^-(t) \right ] \; .
\ee
Substituting this into Eqs. (16) for the case of a single atom, we employ
the properties of operators of spin $1/2$,
$$
S^- S^- = S^+ S^+ = 0 \; , \qquad S^z S^z = \frac{1}{4}\; , \qquad
S^- S^z = \frac{1}{2}\; S^- \; , \qquad S^z S^- = -\;\frac{1}{2}\; S^- \; ,
$$
$$
S^+ S^z =-\; \frac{1}{2}\; S^+ \; , \qquad S^z S^+ = \frac{1}{2}\; S^+ \; ,
\qquad S^- S^+ = \frac{1}{2} -S^z \; , \qquad S^+ S^- =
\frac{1}{2} + S^z \; .
$$
Then we come to the equations for a single atom
\be
\label{22}
\frac{dS^-}{dt} = - i ( \om_0 -\dlt_L - i\gm_0) S^- +
\frac{\bd^2}{|\bd|^2}\; (\gm_0 + i\dlt_L) S^+ \; , \qquad
\frac{dS^z}{dt} = -2\gm_0\left ( \frac{1}{2} + S^z \right ) \; ,
\ee
in which the notation  for the natural width
\be
\label{23}
\gm_0 \equiv \frac{2}{3}\; |\bd|^2 k_0^3
\ee
and the Lamb shift
\be
\label{24}
\dlt_L \equiv \frac{\gm_0}{2\pi} \; \ln \left (
\frac{mc^2}{\hbar\om_0}\right )
\ee
are introduced. The solutions to Eqs. (22), keeping in mind that
$\gm_0\ll\om_0$ and $\dlt_L\ll\om_0$, are
$$
S^-(t) = S^-(0)\exp\left\{ -i(\om_0 -\dlt_L) t -\gm_0 t\right \} \; ,
$$
$$
S^z(t) = -\; \frac{1}{2} + \left [ \frac{1}{2} + S^z(0)\right  ] \;
\exp(-2\gm_0 t) \; .
$$

Thus, the existence of self-action leads to the appearance of attenuation
in the dynamics of the spin operators and to the Lamb frequency shift. The
latter can always be included in the definition of the transition frequency
$\om_0$. Taking into consideration the attenuation, one usually generalizes
the equations of motion by including $\gm_2$, instead of $\gm_0$, for
$S_i^-$ and inserting $\gm_1$, instead of $2\gm_0$, for $S_i^z$.

For a system of $N$ radiating atoms, the vector potential (15) can be
presented as a sum
\be
\label{25}
\bA = \bA_{vac} +\bA_{rad} +\bA_{dip} + \bA_{mat} \; .
\ee
Here $\bA_{vac}$ is caused by vacuum fluctuations. The vector potential
\be
\label{26}
\bA_{rad}(\br,t) = \sum_j \frac{2}{3c|\br-\br_j|} \; \bJ_j \left ( t - \;
\frac{|\br-\br_j|}{c}\right )
\ee
is due to the spherical part of the potential (15), produced by radiating
atoms, which in addition produce the dipolar part
\be
\label{27}
A_{dip}^\al(\br,t) = - \sum_j \sum_\bt \int
\frac{D_{\al\bt}(\br'-\br_j)}{4\pi c|\br-\br'|} \; J_j^\bt \left ( t -\;
\frac{|\br-\br'|}{c}\right ) \; d\br' \; .
\ee
Finally, the action of matter, filling the cavity, creates the vector
potential
\be
\label{28}
A_{mat}^\al(\br,t) = \sum_\bt \int
\frac{\dlt_{\al\bt}(\br'-\br'')}{c|\br-\br'|} \; j_{mat}^\bt\left (\br'',
t-\; \frac{|\br-\br'|}{c}\right )\; d\br' d\br'' \; .
\ee

Let us combine the vacuum, dipole, and matter vector potentials into the
sum
\be
\label{29}
\xi(\br,t) \equiv 2k_0\bd \cdot\left ( \bA_{vac} + \bA_{dip} + \bA_{mat}
\right ) \; ,
\ee
which describes local field fluctuations. Owing to the local nature of the
fluctuating field (29), it can be treated as a random variable. Contrary
to this, the radiation potential (26) is  of long-range nature and can be
responsible for collective effects. The existence of two types of variables,
acting on different spatial scales, makes it possible to employ the scale
separation approach [5,47--50]. Then $\bA_{rad}$ and $\xi$ are considered
as different operators. Since $\bA_{rad}$ is expressed through the spin
operators, we may treat the set $S=\{ \bS_j|\; j=1,2,\ldots,N\}$ of these
operators as approximately commuting with $\xi$. Thus, the total set of
operators consists of two types of the operators, $S$ and $\xi$. For any
operator function $\hat F=\hat F(S,\xi)$, we may introduce two kinds of
averages. One is the average over the spin variables,
\be
\label{30}
<\hat F>\; \equiv {\rm Tr}_S\hat\rho \hat F(S,\xi) \; ,
\ee
with the trace over spins, and $\hat\rho$ being a statistical operator.
Another average is defined as
\be
\label{31}
\ll \hat F\gg \; \equiv {\rm Tr}_\xi\hat\rho \hat F(S,\xi) \; ,
\ee
with the trace over the stochastic field (29).

For the spin averaging (30), we may use the decoupling
\be
\label{32}
< S_i^\al S_j^\bt>\; = \; < S_i^\al>< S_j^\bt> \qquad (i\neq j) \; ,
\ee
keeping in mind the long-range nature of $\bA_{rad}$. This reminds us
the mean-field approximation, which is valid for long-range forces [51].
However, there is a principal difference between the mean-field
approximation and the decoupling (32). The latter involves only the
spin degrees of freedom, not touching the stochastic variables $\xi$,
which are responsible for quantum effects. Employing a seemingly
semiclassical form (32), at the same time preserving quantum features,
associated with stochastic field $\xi$, is close to the method of
stochastic quantization used in quantum field theory [52]. Therefore
the decoupling (32) can be called the {\it stochastic mean-field
approximation} or the {\it quantum mean-field approximation}.

Let us now average the operator equations (16) over the spin variables,
according to Eq. (30), with employing the following notation. The {\it
transition function}
\be
\label{33}
u_i(t) \equiv u(\br_i,t) \equiv 2<S_i^-(t)>
\ee
describes the local dipole transitions. The local characteristic of
coherence is the {\it coherence intensity}
\be
\label{34}
w_i(t) \equiv w(\br_i,t) \equiv \frac{2}{n_0} \;
\sum_{<j>}^{n_0} \left [ <S_i^+(t) S_j^-(t)> + <S_j^+(t) S_i^-(t)>
\right ] \; ,
\ee
in which the sum is over the nearest neighbours and $n_0$ is the number
of the latter. Finally, the local {\it population difference} is
\be
\label{35}
s_i(t) \equiv s(\br_i,t) \equiv 2< S_i^z(t)>\; .
\ee

It is also convenient to pass to the continuous spatial representation,
replacing the sums by the integrals as
\be
\label{36}
\sum_{j=1}^N \Longrightarrow \rho \int d\br \qquad \left ( \rho \equiv \frac{N}{V}
\right ) \; .
\ee

Let us introduce the notation for the effective field potential acting
on atoms,
\be
\label{37}
f(\br,t) = f_0(\br,t) + f_{rad}(\br,t) + \xi(\br,t) \; ,
\ee
which consists of the part
\be
\label{38}
f_0(\br,t) \equiv -2 i \bd\cdot \bE_0(\br,t) \; ,
\ee
due to an external electric field $\bE_0$, of the term
\be
\label{39}
f_{rad}(\br,t) \equiv 2k_0<\bd\cdot \bA_{rad}(\br,t)> \; ,
\ee
caused by the radiating atoms, and of the fluctuating random field (29).
The radiation potential (39), taking account of Eq. (26), can be written
as
$$
f_{rad}(\br,t) = -i\gm_0\rho \int \left [ G(\br-\br',t) u(\br',t) -
\; \frac{\bd^2}{|\bd|^2} \; G^*(\br-\br',t) u^*(\br',t) \right ]\;
d\br' \; ,
$$
with the transfer kernel
$$
G(\br,t) \equiv \frac{\exp(ik_0 r)}{k_0 r}\; \Theta(ct-r) \; .
$$

In this way, from Eqs. (16), we obtain for the local functions (33) to
(35) the evolution equations
$$
\frac{\prt u}{\prt t} = - (i\om_0 +\gm_2) u + fs \; , \qquad
\frac{\prt w}{\prt t} = -2\gm_2 w + (u^* f + f^* u) s \; ,
$$
\be
\label{40}
\frac{\prt s}{\prt t} = -\; \frac{1}{2}\; \left ( u^* f + f^* u \right )
-\gm_1 ( s-\zeta) \; ,
\ee
where $\gm_1$ and $\gm_2$ are the longitudinal and transverse attenuation
rates and $\zeta$ is the stationary pumping parameter. These stochastic
differential equations are the basic equations to be used in what follows
for describing the spatio-temporal evolution in lasers.

\section{Turbulent Photon Filamentation}

The external field $\bE_0$ is here the seed field
\be
\label{41}
\bE_0(\br,t) = \frac{1}{2}\; \bE_1 e^{i(kz-\om t)} +
\frac{1}{2}\; \bE_1^* e^{-i(kz-\om t)} \; ,
\ee
selecting a longitudinal mode of frequency $\om=kc$, but imposing no
constraints on possible transverse modes. The propagation of the field
(41) is along the axis $z$, which is the axis of a cylindrical laser
cavity. The frequency $\om$ has to be in resonance with the atomic
transition frequency $\om_0$, so that the detuning be small,
\be
\label{42}
\frac{|\Dlt|}{\om_0} \ll 1 \qquad (\Dlt\equiv \om - \om_0) \; .
\ee
The wavelength $\lbd\equiv 2\pi c/\om$ is usually much smaller than the
effective laser radius $R$ and length $L$,
\be
\label{43}
\frac{\lbd}{R}\ll 1 \; , \qquad \frac{\lbd}{L} \ll 1 \; .
\ee
The seed field (41) is very weak, such that
\be
\label{44}
\frac{|\nu_1|}{\om_0} \ll 1 \qquad (\nu_1 \equiv\bd\cdot\bE_1) \; .
\ee

Keeping in mind the possibility of arising filamentary structures,
we should look for the solution of Eqs. (40) in the form of the modal
superposition
$$
u(\br,t) = \sum_{n=1}^{N_f} u_n(\br_\perp,t) e^{ikz} \; , \qquad
w(\br,t) = \sum_{n=1}^{N_f} w_n(\br_\perp,t) \; ,
$$
\be
\label{45}
s(\br,t) = \sum_{n=1}^{N_f} s_n(\br_\perp,t) \; ,
\ee
in which $N_f$ is the number of filaments and
$r_\perp\equiv\sqrt{x^2+y^2}$. Note that the representation (45)
is rather general and includes as well the case of no filamentatory
structure, when $N_f=1$. The number of filaments $N_f$ will be defined
later in a self-consistent way. From the point of view of quantum field
theory,  the appearance of spatial structures corresponds to the existence
of nonuniform field vacuum [53,54]. In the case when different filaments
are not correlated with each other, one has
\be
\label{46}
u_m(r_\perp,t) s_n(r_\perp,t) = \dlt_{mn} u_n(r_\perp,t)
s_n(r_\perp,t) \; .
\ee
This condition is typical of the turbulent regime, when the filaments
are not mutually correlated [36--42]. Since filaments do not interact
with each other, their location in the transverse cross-section is
random. The radiation inside each filament is mainly concentrated
along the filament axis, fading away at the distance much larger than
the filament radius. In general, there can simultaneously exist the
filaments of different radii.

Let us consider an $n$-th filament. And let the radiation in this
filament be an order of magnitude weaker at the distance $R_n$
from its axis than at the latter, so that the intensity function
$w_n(R_n,t)$ is an order of magnitude smaller at $r_\perp=R_n$
than $w_n(0,t)$ at $r_\perp=0$. The effective radius of the filament,
$r_n$, can be defined by the averaging relation
\be
\label{47}
\frac{2}{R_n^2} \; \int_0^{R_n} w_n(r_\perp,t) r_\perp\; dr_\perp=
w_n(r_n,t) \; .
\ee
If the profile of the intensity function $w_n$ is of normal law, that
is,
\be
\label{48}
w_n(r_\perp,t) = w_n(0,t)\exp\left ( -\; \frac{r_\perp^2}{2r_n^2}
\right ) \; ,
\ee
where the filament radius $r_n$ plays the role of the standard deviation,
then the relation (47) yields
\be
\label{49}
r_n = \frac{R_n}{(4e)^{1/4}} = 0.55 \; R_n \; .
\ee

All radiation of a filament, with an effective radius $r_n$, is
concentrated inside the enveloping cylinder of radius $R_n$. Let us
define the averaged functions
$$
u(t) \equiv \frac{1}{V_n}\; \int_{V_n} u_n(r_\perp,t) \; d\br \; , \qquad
w(t) \equiv \frac{1}{V_n}\; \int_{V_n} w_n(r_\perp,t)\; d\br \; ,
$$
\be
\label{50}
s(t) \equiv \frac{1}{V_n} \; \int_{V_n} s_n(r_\perp,t)\; d\br \; ,
\ee
where the averaging is over the enveloping cylinder of volume
$V_n\equiv\pi R_n^2L$, and the enumeration index $n$ for the left-hand
side functions is omitted in order to simplify the notation.

For what follows, we shall need the definition of the coupling functions
\be
\label{51}
\al(t) \equiv \gm_0\rho \int_{V_n} \Theta(ct-r) \;
\frac{\sin(k_0r-kz)}{k_0 r}\; d\br
\ee
and
\be
\label{52}
\bt(t) \equiv \gm_0\rho \int_{V_n} \Theta(ct-r) \;
\frac{\cos(k_0r-kz)}{k_0 r}\; d\br \; .
\ee
Introduce also the averaged stochastic field
\be
\label{53}
\xi(t) \equiv \frac{1}{V_n} \; \int \xi(\br,t) e^{-ikz}\; d\br
\ee
and the field potential
\be
\label{54}
f_1(t) \equiv -i\bd\cdot \bE_1 e^{-i\om t} + \xi(t) \; .
\ee
Then, substituting the representation (45) into the evolution
equations (40) and averaging according to Eq. (50), we come to the
equations
$$
\frac{du}{dt} = - i(\om_0+\bt s)u - (\gm_2 -\al s)u + f_1 s \; ,
\qquad \frac{dw}{dt} = - 2(\gm_2 -\al s)w +
(u^* f_1 + f_1^* u)s  \; ,
$$
\be
\label{55}
\frac{ds}{dt} = -\al w  -\;\frac{1}{2} (u^* f_1 + f_1^* u)s -
\gm_1(s-\zeta) \; ,
\ee
describing the dynamics of each of the filaments.

In this way, the representation (45), together with the averaging
procedure (50), have made it possible to pass from the equations (40)
in partial derivatives to the ordinary differential equations (55).
Following further the scale separation approach [5,47--50], we can
more simplify Eqs. (55) by taking into account the existence of small
parameters related to the standard situation, when the attenuation
rates are essentially smaller than the transition frequency,
\be
\label{56}
\frac{\gm_0}{\om_0} \ll 1 \; , \qquad \frac{\gm_1}{\om_0}\ll 1 \; ,
\qquad \frac{\gm_2}{\om_0} \ll 1 \; .
\ee
This tells us that the function $u$ in Eqs. (55) is fast, as
compared to the slow functions $w$ and $s$, which are the temporal
quasi-invariants of motion. The collective width
\be
\label{57}
\Gamma \equiv \gm_2 - \al s \; ,
\ee
collective frequency
\be
\label{58}
\Om \equiv \om_0 +\bt s \; ,
\ee
and effective detuning
\be
\label{59}
\dlt \equiv \om - \Omega = \Dlt -\bt s
\ee
are also slow functions in time, as compared to the fast function $u$.
The latter can be found from the first of Eqs. (55), which gives
$$
u =\left ( u_0 - \; \frac{\nu_1 s}{\dlt + i\Gamma}\right )\;
e^{-(i\Om+\Gamma)t} + \frac{\nu_1 s}{\dlt+i\Gamma}\; e^{-i\om t} +
$$
\be
\label{60}
+ s \int_0^t \xi(t') e^{-(i\Omega+\Gamma)(t-t')}\; dt' \; .
\ee
Without the loss of generality, it is possible to choose the phase of the
external field $\bE_1$ so that $u_0^*\bd\cdot\bE_1$ be real, that is,
\be
\label{61}
u_0^* \nu_1 = u_0 \nu_1^*\; , \qquad u_0 \equiv u(0) \; .
\ee

The solution (60) has to be substituted into the second and third of Eqs.
(55), which are the equations for the slow functions. Then the right-hand
sides of these equations are to be averaged over time and over the
stochastic variables according to the rule
$$
\lim_{\tau\ra\infty}\; \frac{1}{\tau} \; \int_0^\tau \ll \ldots
\gg \; dt \; ,
$$
keeping fixed all temporal quasi-invariants. The stochastic variable $\xi$,
describing local fluctuations, is assumed to be such that
\be
\label{62}
\ll \xi(t) \gg \; = 0 \; .
\ee

Define the {\it effective attenuation}
\be
\label{63}
\tilde\Gamma \equiv \gm_3 + \frac{|\nu_1|^2\Gamma}{\dlt^2+\Gamma^2}\;
\left ( 1 - e^{-\Gamma t} \right ) \; ,
\ee
in which $|\dlt|<|\Gamma|$ and the first term is caused by the quantum
local fluctuations resulting in the {\it quantum attenuation}
\be
\label{64}
\gm_3 \equiv {\rm Re}\; \lim_{\tau\ra\infty} \; \frac{1}{\tau} \;
\int_0^\tau dt \int_0^t \ll \xi^*(t)\xi(t')\gg \;
e^{-(i\Omega+\Gamma)(t-t')} \;  dt' \; .
\ee
Thus we obtain the equations for the guiding centers
\be
\label{65}
\frac{dw}{dt} = -2(\gm_2-\al s) w + 2\tilde\Gamma s^2 \; , \qquad
\frac{ds}{dt} = -\al w - \tilde\Gamma s - \gm_1(s-\zeta) \; .
\ee
Here the fast field fluctuations have been averaged out, and only the
slow dynamics of filaments is left.

\section{Transition Dipolar Waves}

An essential part of the quantum local fluctuations, leading to the
quantum attenuation (64), is due to the dipolar vector potential (27)
entering the local field (29). In order to better understand the
physical origin of these fluctuations, let us consider the dipolar
part of the local field (29) having the form
\be
\label{66}
2k_0\bd\cdot \bA_{dip}(\br_i,t) = i\sum_{j(\neq i)} \left [
b_{ij}^* S_j^-(t) - c_{ij} S_j^+(t) \right ] \; ,
\ee
where we use the notation
$$
b_{ij}^* \equiv \frac{k_0^2}{2\pi} \; \sum_{\al\bt} d^\al \left (
D_{ij}^{\al\bt} d^\bt\right )^* \; , \qquad
c_{ij} \equiv \frac{k_0^2}{2\pi} \; \sum_{\al\bt} d^\al
D_{ij}^{\al\bt} d^\bt  \; ,
$$
$$
D_{ij}^{\al\bt} \equiv \int \Theta(ct-|\br_i-\br'|) \;
\frac{D_{\al\bt}(\br'-\br_j)}{|\br_i-\br'|} \; \exp\left (
-ik_0 |\br_i-\br'|\right )\; d\br' \; .
$$
Leaving in the pseudospin equations (16) only the terms related to the
dipolar part of the vector potential, we have
$$
\frac{dS_i^-}{dt} = -i\om_0 S_i^- + iS_i^z \sum_{j(\neq i)} \left (
b_{ij}^* S_j^- - c_{ij} S_j^+\right ) \; ,
$$
\be
\label{67}
\frac{dS_i^z}{dt} = - \; \frac{i}{2}\; S_i^+ \sum_{j(\neq i)} \left (
b_{ij}^* S_j^- - c_{ij} S_j^+\right ) \; .
\ee
These equations describe the pseudospin fluctuations, which can be
characterized by the deviations
\be
\label{68}
\dlt S_i^- = S_i^- - \; < S_i^-> \; , \qquad
\dlt S_i^z = S_i^z - \; < S_i^z>
\ee
from the corresponding average values. Linearizing Eqs. (67) with respect
to the deviations (68), under the condition $<S_i^->=0$, shows that
$$
S_i^- = \dlt S_i^- \; , \qquad S_i^z = const \; .
$$
Employing the Fourier transforms for the pseudospin operators
\be
\label{69}
S_j^- = \sum_k\; S_k^- \; \exp\left ( i\bk\cdot \br_j\right ) \; ,
\qquad S_k^- = \frac{1}{N} \sum_j\; S_j^- \; \exp\left (
-i\bk\cdot \br_j\right ) \;
\ee
and, similarly, for the coefficients
$$
b_{ij} = \sum_k\; b_k \; \exp\left ( i\bk\cdot \br_j\right ) \; ,
\qquad c_{ij} = \sum_k\; c_k \; \exp\left ( i\bk\cdot \br_j
\right ) \; ,
$$
where $\br_{ij}\equiv\br_i-\br_j$, we get the equations for $S_k^-$ and
$S_k^+$ which read
\be
\label{70}
\frac{dS_k^-}{dt} = -i\mu_k S_k^- - i\lbd_k S_k^+ \; , \qquad
\frac{dS_k^+}{dt} = i\mu_k^* S_k^+ + i\lbd_k^* S_k^- \; ,
\ee
where
\be
\label{71}
\mu_k \equiv \om_0 - b_k^*\; < S_i^z> \; , \qquad
\lbd_k \equiv c_k\; <S_i^z> \; .
\ee
The solution to Eqs. (70) has the form
\be
\label{72}
S_k^- = u_k e^{-i\om_kt} + v_k^* e^{i\om_kt} \; ,
\ee
with the spectrum
\be
\label{73}
\om_k = \sqrt{|\mu_k|^2-|\lbd_k|^2} \; .
\ee
This means that the dipolar part of the vector potential generates local
fluctuations realized as a kind of transition dipolar waves. Such waves
are analogous to the dipolar spin waves in magnets [55]. The spectrum
(73) is always positive, since $|b_k|^2=|c_k|^2\ll\om_0$. This implies
that the transition dipolar waves are stable. In the long-wave limit
$k\ra 0$, one has $b_0=c_0$, because of which
\be
\label{74}
\lim_{k\ra 0} \om_k = \om_0 \; .
\ee

The transition dipolar waves are responsible only for local fluctuations,
but they do not participate in collective effects which are related to
the coupling functions (51) and (52). These functions are zero at the
initial time $t=0$, though they grow very fast. Collective effects come
to play after the interaction time $\tau_{int}=a/c$, where $a$ is the
nearest-neighbour distance and $c$ light velocity. For $a\sim 10^{-8}$
cm and $c\sim 10^{10}$ cm/s, this time is very short, $\tau_{int}\sim
10^{-18}$ s. After this time, the coupling function (51) quickly reaches
its maximal value $g\gm_2$ and (52) grows to $g'\gm_2$. Here the coupling
parameters
\be
\label{75}
g\equiv \rho\; \frac{\gm_0}{\gm_2} \; \int_{V_n}
\frac{\sin(k_0r-kz)}{k_0r}\; d\br
\ee
and, respectively,
\be
\label{76}
g' \equiv \rho\; \frac{\gm_0}{\gm_2} \; \int_{V_n}
\frac{\cos(k_0r-kz)}{k_0r}\; d\br
\ee
are introduced. Recall that for each filament there are its own coupling
parameters, whose values depend on the filament characteristics and the
number of atoms in the filament. Strictly speaking, collective effects
appear already for two atoms [56,57]. However noticeable radiation
coherence develops only when a large number of atoms is involved.

For $t\gg\tau_{int}$, the collective width (57) becomes
\be
\label{77}
\Gamma =\gm_2(1 - gs)
\ee
and the collective frequency (58) is
\be
\label{78}
\Om= \om_0 + g'\gm_2 s \; .
\ee
When the seed field is negligibly small, so that $|\nu_1|\ll\gm_2$, then
the effective attenuation (63) simplifies to
\be
\label{79}
\tilde\Gm\simeq \gm_3 \; .
\ee
Therefore, for times after $\tau_{int}$, the evolution equations (65)
rearrange to the equations
\be
\label{80}
\frac{dw}{dt} = -2\gm_2 ( 1 - gs) w + 2\gm_3 s^2 \; , \qquad
\frac{ds}{dt} = -g\gm_2 w - \gm_3 s  -\gm_1 (s -\zeta) \; ,
\ee
characterizing the filament dynamics. Recall that Eqs. (80) describe
the slow dynamics, since the uncorrelated fast oscillations have been
averaged out in the process of derivation of Eqs. (80). The random fast
fluctuations are of less physical importance, being, in addition, much
smaller than the correlated slow functions [36--42].

\section{Temporal Dynamics of Filaments}

Equations (80), describing the filament dynamics, are to be complemented
by the initial conditions $w_0=w(0)$ and $s_0=s(0)$. The first dynamic
stage in the interval $0<t<\tau_{int}$ can be ignored because of a very
short interaction time $\tau_{int}$. In order that at the initial time
$t=0$ the coherence intensity would be an increasing function of time,
such that $dw/dt>0$, it is necessary and sufficient that the inequality
\be
\label{81}
\gm_2(gs_0-1)w_0 + \gm_3 s_0^2 > 0
\ee
be valid, which is a criterion of laser generation. Since the second
term in Eq. (81) is always non-negative, a sufficient condition for
laser generation is $gs_0>1$.

A necessary condition for the formation of filaments is the appearance
of coherence between radiating atoms. At the beginning, there exists
a quantum stage of spontaneously radiating atoms, before coherence
develops to a noticeable amount. The incoherent quantum stage lasts
till the crossover time $t_c$, after which coherent effects become
dominant, and filaments are being formed. During the quantum stage
$0<t<t_c$, the development of coherence is caused by the quantum
fluctuations related to the term $2\gm_3 s^2$ in the first of Eqs.
(80). At this quantum stage, the system is yet uniform. Filaments are
well formed after the crossover time $t_c$. The formation of filaments
inside a uniform atomic system, due to local quantum fluctuations, is
somewhat analogous to the formation of galaxies from a uniform matter,
due to local fluctuations [58], or to the stratification in quantum
systems [59,60]. But the principal feature of laser filamentation is
that the atomic matter in the laser cavity does not stratify as such.
Nonuniformity happens for the distribution of radiating atoms, that
is, it occurs on the level of photons, but not atoms themselves. This
is why the filamentation in lasers can be called the {\it photon
filamentation}.

To estimate the crossover time $t_c$, we need, first, to consider
the quantum stage of evolution, when there is no yet self-organized
coherence. Assuming that at the initial time there is no coherence
imposed by external fields, so that $w_0=0$, we get the equations for
the quantum stage,
\be
\label{82}
\frac{dw}{dt} = 2\gm_3 s^2 \; , \qquad
\frac{ds}{dt} = -(\gm_1+\gm_3) s + \gm_1 \zeta \; .
\ee
The solution for the population difference is
$$
s=\left ( s_0 - \; \frac{\gm_1\zeta}{\gm_1+\gm_3}\right ) \;
\exp\{ -(\gm_1+\gm_3)\; t\} + \frac{\gm_1\zeta}{\gm_1+\gm_3} \; .
$$
At short time, when $(\gm_1+\gm_3)t\ll 1$, this yields
\be
\label{83}
s\simeq s_0 - \left ( s_0 -\; \frac{\gm_1\zeta}{\gm_1+\gm_3}\right ) \;
(\gm_1+\gm_3)\; t + \frac{1}{2}
\left ( s_0 -\; \frac{\gm_1\zeta}{\gm_1+\gm_3}\right ) \;
(\gm_1+\gm_3)^2\; t^2 \; .
\ee
The latter form, since usually $\gm_1<\gm_3$, is valid if $\gm_3t_c\ll 1$.
At this quantum stage, the solution for the coherence intensity, given by
the first of Eqs. (82), essentially depends on the initial value $s_0$ of
the population difference. For an arbitrary $s_0$. we have
$$
w\simeq 2\gm_3 s_0^2 t - 2\gm_3s_0(\gm_1+\gm_3)\left ( s_0 -\;
\frac{\gm_1\zeta}{\gm_1+\gm_3}\right )\; t^2 +
$$
\be
\label{84}
 + \frac{2}{3}\; \gm_3 (\gm_1+\gm_3)^2 \left ( s_0 - \;
\frac{\gm_1\zeta}{\gm_1+\gm_3}\right ) \left ( 2s_0 - \;
\frac{\gm_1\zeta}{\gm_1+\gm_3}\right )\; t^3 \; .
\ee
When $s_0\neq 0$, then the leading term in the coherence intensity is
linear in time,
$$
w\simeq 2\gm_3 s_0^2 t \qquad (s_0\neq 0) \; .
$$
But if $s_0=0$, then Eq. (84) gives the cubic dependence
$$
w\simeq \frac{2}{3}\; \gm_1^2 \gm_3 \zeta^2 t^3 \qquad (s_0=0) \; .
$$

During the quantum stage, the evolution of coherence is mainly due to the
term $2\gm_3s^2$ in the first of Eqs. (80). But at the following coherent
stage, the term $2\gm_2(gs-1)w$ in this equation becomes dominant.
Therefore, the crossover time $t_c$ can be defined as that one, where
the two terms of different nature coincide,
\be
\label{85}
\gm_2(gs-1) w = \gm_3 s^2 \qquad (t =t_c) \; .
\ee

Note that under the initial condition $s_0=0$, no noticeable coherence can
evolve. This follows from Eq. (85) which, together with Eqs. (83) and (84),
yields $t_c\sim T_2\equiv 1/\gm_2$. The rising of coherence should occur
during the crossover time $t_c\ll T_2$.

If $s_0\neq 0$, then Eq. (85) gives the crossover time
\be
\label{86}
t_c =\frac{s_0/2}{\gm_1(s_0-\zeta)+\gm_2(gs_0-1)s_0+\gm_3s_0} \; .
\ee
Usually, one has $\gm_1\ll\gm_2\sim\gm_3$, and, as is discussed above,
one should have $gs_0>1$ in order that collective effects could be
important. Then the crossover time (86) can be reduced to
\be
\label{87}
t_c = \frac{T_2}{2(gs_0-1)} \; .
\ee
For $gs_0\gg 1$, it is evident that $t_c\ll T_2$. At the crossover time
(87), solutions (83) and (84) can be approximated as
\be
\label{88}
w(t_c)\simeq 2\gm_3 t_c s_0^2 \; , \qquad s(t_c)\simeq s_0 \; .
\ee

After the crossover time $t_c$, coherent effects become important. For
the coupling parameter $g\gg 1$, keeping in mind that $\gm_2\sim\gm_3$,
one has $g\gm_2\gg\gm_3$. In the standard situation, $\gm_1\ll\gm_2$,
because of which in the temporal interval $t_c<t\ll T_1\equiv1/\gm_1$
equations (80) can be simplified to the evolution equations
\be
\label{89}
\frac{dw}{dt} = -2\gm_2(1-gs)w\; , \qquad
\frac{ds}{dt} = -g\gm_2 w\; ,
\ee
describing the transient coherent stage of filament dynamics. These
equations can be solved exactly, resulting in the solution
\be
\label{90}
w=\left ( \frac{\gm_p}{g\gm_2}\right )^2\; {\rm sech}^2 \left (
\frac{t-t_0}{\tau_p} \right ) \; , \qquad
s= \frac{1}{g}\; - \; \frac{\gm_p}{g\gm_2}\; {\rm tanh}\left (
\frac{t-t_0}{\tau_p} \right ) \; ,
\ee
whose integration parameters are obtained from the initial conditions
(88). The {\it pulse width} $\gm_p$ is defined by the relations
\be
\label{91}
\gm_p^2\equiv \gm_g^2 + 2(g\gm_2)^2 \gm_3 t_c s_0^2 \; , \qquad
\gm_g\equiv (gs_0-1)\gm_2 \; , \qquad \gm_p\tau_p \equiv 1 \; .
\ee
The {\it pulse time}, taking into account that $\gm_3t_c\ll 1$, reads
as
\be
\label{92}
\tau_p = \frac{T_2}{gs_0-1}\left [ 1 - \;
\frac{\gm_3t_cg^2s_0^2}{(gs_0-1)^2} \right ] \; .
\ee
And the {\it delay time} is
\be
\label{93}
t_0 = t_c + \frac{\tau_p}{2}\; \ln\left |
\frac{\gm_p+\gm_g}{\gm_p-\gm_g}\right | \; .
\ee
Recall that solution (90) is valid only if the pulse time $\tau_p$
is much less than $T_1$. When $\tau_p$ is of order or much longer than
$T_1$, as in many experiments [36--42], than one needs to deal with the
total Eqs. (80).

In view of the fact that $\gm_3t_c\ll 1$, the pulse width can be written
as
$$
\gm_p \simeq (gs_0-1)\gm_2 + \frac{g^2\gm_2\gm_3t_cs_0^2}{gs_0-1} \; ,
$$
where it is assumed that $gs_0>1$. Then the delay time (93) becomes
\be
\label{94}
t_0 = t_c + \frac{\tau_p}{2}\; \ln\left |
\frac{2(gs_0-1)^2}{g^2\gm_3t_cs_0^2}\right | \; .
\ee
The latter, under strong coupling, when $gs_0\gg 1$ and
$\tau_p\simeq 2t_c$, rearranges to
$$
t_0 \simeq t_c + t_c\ln\left | \frac{2}{\gm_3t_c}\right | \; ,
$$
with $t_c\simeq T_2/2gs_0$.

At the final dynamic stage, when $t\gg T_1$, we have to consider all terms
in Eqs. (80). These can be written in the form
\be
\label{95}
\frac{dw}{dt}=v_1 \; , \qquad \frac{ds}{dt}=v_2 \; ,
\ee
with the notation
$$
v_1 = -2\gm_2(1-gs) w + 2\gm_3 s^2\; , \qquad
v_2=-g\gm_2 w -\gm_3 s -\gm_1(s-\zeta) \; .
$$
The stability properties of the solutions to Eq. (95) are characterized
by the Jacobian matrix $\hat J(t)=[J_{ij}(t)]$, with the elements
$$
J_{11}\equiv \frac{\prt v_1}{\prt w} = 2\gm_2(gs-1) \; , \qquad
J_{12}\equiv \frac{\prt v_1}{\prt s} = 2\gm_2 gw + 4\gm_3 s\; ,
$$
\be
\label{96}
J_{21} \equiv \frac{\prt v_2}{\prt w} = - g\gm_2 \; , \qquad
J_{22} \equiv \frac{\prt v_2}{\prt s} = -\gm_1 -\gm_3 \; .
\ee

Considering the stationary solutions to Eqs. (95), given by the equations
$v_1=v_2=0$, we will present only those of them that correspond to stable
fixed points. It is possible to separate out three cases, depending on the
values of the coupling $g$ and pumping parameter $\zeta$.

When $g\zeta\ll-1$, then the stationary solutions
\be
\label{97}
w^* \simeq \frac{\gm_3|\zeta|}{\gm_2|g|} \; , \qquad
s^* \simeq \zeta\left ( 1  -\; \frac{\gm_3}{\gm_1|g\zeta|} \right )\; ,
\ee
present a stable node, since the eigenvalues of the Jacobian matrix, that
is, the characteristic exponents, are
$$
J_1 \simeq -\gm_1 -\gm_3\; , \qquad J_2 \simeq -2\gm_2|g\zeta| \; .
$$
Then the sole transient coherent pulse occurs, after which the radiation
intensity diminishes to a value proportional to $w^*$.

For $|g\zeta|\ll 1$, the stationary solutions
\be
\label{98}
w^* \simeq \left ( \frac{\gm_1\zeta}{\gm_1+\gm_3}\right )^2
\frac{\gm_3}{\gm_2}\left [ 1 +
\frac{\gm_1(\gm_1-\gm_3)g\zeta}{(\gm_1+\gm_3)^2} \right ] \; , \qquad
s^* \simeq \frac{\gm_1\zeta}{\gm_1+\gm_3}\left ( 1  -\;
\frac{\gm_1\gm_3g\zeta}{(\gm_1+\gm_3)^2}\right )
\ee
also corresponds to a stable node, with the characteristic exponents
$$
J_1 \simeq -\gm_1 -\gm_3 \; , \qquad J_2 \simeq -2\gm_2 \; .
$$
Hence again, only a sole transient pulse can appear.

Finally, when both the coupling and pumping are strong, so that
$g\zeta\gg 1$, the stationary solutions
\be
\label{99}
w^* \simeq \frac{\gm_1\zeta}{\gm_2 g} \; , \qquad s^* \simeq \frac{1}{g}
\left ( 1  -\; \frac{\gm_3}{\gm_1g\zeta}\right )
\ee
represent a stable focus, with the characteristic exponents
$$
J_{1,2} \simeq -\; \frac{1}{2}\; (\gm_1 +\gm_3) \pm i\om_\infty \; ,
$$
where the effective asymptotic frequency
$$
\om_\infty \equiv \sqrt{2g\zeta\gm_1\gm_2} \; .
$$
This means that the strong pumping supports the pulsing regime of
radiation, when a set of bursts arise, being, at long times, separated
from each other by the asymptotic period
\be
\label{100}
T_\infty \equiv \frac{2\pi}{\om_\infty} = \pi\;
\sqrt{\frac{2T_1T_2}{g\zeta} } \; .
\ee
The pulsing regime of radiation resembles pulsating stationary states
happening in some statistical systems [61].

In this way, the flashing dynamics of each radiating filament depends
on the related coupling parameter $g$ and the level of stationary
pumping defined by the pumping parameter $\zeta$. In general, there
are the following stages of evolution. The {\it interaction stage}
$0<t<\tau_{int}$, when each atom radiates independently, before the
signal reaches its nearest neighbours. This stage is very short and
usually can be neglected. The {\it quantum stage} $\tau_{int}<t<t_c$,
when the radiation dynamics is mainly governed by local quantum
fluctuations. After the crossover time $t_c$, coherent collective
effects become important, signifying the presence of the {\it coherent
stage}. At this latter stage, depending on the value of the pumping
parameter $\zeta$, there occurs either a sole coherent pulse or a
series of coherent bursts. If a stationary pumping is absent, then
the coherent stage lasts in the interval $t_c<t<t_0+\tau_p$, and for
$t\gg t_0$ it changes to the {\it relaxation stage}, when the coherence
intensity exponentially diminishes to the low level (97).

\section{Spatial Structure of Filaments}

Filaments are randomly distributed in the transverse cross-section
of the laser cavity, evolving in space and time independently of each
other. The characteristics of each filament essentially depend on the
value of the related coupling parameter (75). For cylindric symmetry
the latter can be presented in the form
\be
\label{101}
g = 2\pi \rho \; \frac{\gm_0}{\gm_2} \;
\int_0^{R_n} r_\perp\; dr_\perp \; \int_{-L/2}^{L/2} \;
\frac{\sin(k_0\sqrt{r_\perp^2+z^2}-kz)}{k_0\sqrt{r_\perp^2+z^2}}\;
dz \; .
\ee
Keeping in mind the resonance condition $k_0\approx k$ and introducing
the variable $x=k(\sqrt{r_\perp^2+z^2}-z)$, we have
\be
\label{102}
g =2\pi\; \frac{\rho\gm_0}{k\gm_2}\; \int_0^{R_n} r_\perp\; dr_\perp\;
\int_{k r_\perp^2/L}^{kL} \; \frac{\sin x}{x}\; dx \; .
\ee
Since, according to the inequality (43), we have $\lbd\ll L$, then the
upper limit $kL$ in the integral (102) can be replaced by $kL\ra\infty$.
This gives
\be
\label{103}
g = 2\pi\; \frac{\rho\gm_0}{k\gm_2} \; \int_0^{R_n} \left [
\frac{\pi}{2}\; - \; {\rm Si}\left ( \frac{k r_\perp^2}{L}\right )
\right ] r_\perp\; dr_\perp \; ,
\ee
with the integral sine
$$
{\rm Si}(x) \equiv \int_0^x \frac{\sin u}{u}\; du = \frac{\pi}{2}
+ \int_\infty^x \frac{\sin u}{u}\; du \; .
$$
Introducing the notation
\be
\label{104}
\vp \equiv \frac{\pi R_n^2}{\lbd L} \; ,
\ee
varying in the interval $0\leq\vp\leq\pi F$ and playing the role of an
effective Fresnel number for a given filament, we transform Eq. (103) to
\be
\label{105}
g(\vp) = \pi \; \frac{\rho\gm_0 L}{k^2\gm_2} \left [ \pi\vp -
\int_0^{2\vp} {\rm Si}(x)\; dx\right ] \; .
\ee

In the same manner, the coupling (76) can be reduced to
\be
\label{106}
g'(\vp) = -\pi\; \frac{\rho\gm_0 L}{k^2\gm_2} \;
\int_0^{2\vp} {\rm Ci}(x)\; dx \; ,
\ee
with the integral cosine
$$
{\rm Ci}(x) \equiv \int_\infty^x \frac{\cos u}{u}\; du \; .
$$

Performing the integration in Eqs. (105) and (106), we may write
$$
g(\vp) = \pi\; \frac{\rho\gm_0 L}{k^2\gm_2}\; \left [ \pi\vp -
2\vp{\rm Si}(2\vp) + 1 - \cos(2\vp)\right ] \; ,
$$
$$
g'(\vp) = \pi\; \frac{\rho\gm_0 L}{k^2\gm_2}\; \left [ \sin(2\vp) -
2\vp{\rm Ci}(2\vp) \right ] \; .
$$
Thus, the coupling parameters are functions of the effective variable
(104), which, in turn, depends on the enveloping radius $R_n$ related
to the effective filament radius $r_n$ by Eq. (49). In general, the
filaments of different radii can arise. However, some of them are more
stable than other, because of which the overwhelming majority of filaments
possess the radii close to a typical value.

The distribution of filaments with respect to their radii and, hence,
the typical radius, can be found by invoking the general method of
probabilistic pattern selection [18,21,22]. Following this approach, we
define the probability distribution
\be
\label{107}
p(\vp,t) = \frac{1}{Z(t)}\; \exp\{-X(\vp,t)\}
\ee
for a filament characterized by the variable $\vp$ at the moment of time
$t$. Here
\be
\label{108}
X(\vp,t) ={\rm Re}\; \int_0^t {\rm Tr}\hat J(\vp,t')\; dt'
\ee
is the {\it expansion exponent}, expressed through the Jacobian matrix
$\hat J$ of the evolution equations, and
$$
Z(t) = \int \exp\{ - X(\vp,t)\} \; d\vp
$$
is the normalizing factor. The expansion exponent (108) defines the {\it
local expansion rate}
\be
\label{109}
\Lambda(\vp,t) \equiv \frac{1}{t}\; X(\vp,t) \; .
\ee
The latter can be represented as the sum of the local Lyapunov exponents
[18,21,22]. The partial sum of only positive Lyapunov exponents defines
the entropy production rate [62,63], hence, the latter does not coincide
with the local expansion rate (109).

Thus, the probability for the appearance of filaments, characterized
by the parameter $\vp$, is given by the probability distribution (107).
As is evident, the most probable is the filament with a typical $\vp$
satisfying the {\it principle of minimal expansion} [18,21,22]
\be
\label{110}
\max_\vp p(\vp,t) \Longleftrightarrow \min_\vp X(\vp,t)
\Longleftrightarrow \min_\vp \Lambda(\vp,t) \; .
\ee
This general principle follows from the minimization of the pattern
information [22] and can be employed for arbitrary dynamical systems.

The problem of turbulent photon filamentation is described by the
evolution equations (80) or (95). The corresponding Jacobian matrix is
given by Eqs. (96) from where
$$
{\rm Tr} \hat J(\vp,t) = -\gm_1 -\gm_3 - 2\gm_2(1-gs) \; ,
$$
with $g=g(\vp)$ and $s=s(t)$. For $t\gg T_1$, the expansion rate (109)
can be presented as
\be
\label{111}
\Lambda(\vp,t) \simeq  -\gm_1 -\gm_3 - 2\gm_2(1-gs^*) \; .
\ee
Using Eqs. (97) to (99), we get
$$
\Lambda(\vp,t) \simeq  -\gm_1 -\gm_3 -2\gm_2 (1 + |g\zeta|)  \qquad
(g\zeta\ll -1) \; ,
$$
$$
\Lambda(\vp,t) \simeq  -\gm_1 -\gm_3 - 2\gm_2\left ( 1- \;
\frac{\gm_1g\zeta}{\gm_1+\gm_3}\right ) \qquad (|g\zeta|\ll 1) \; ,
$$
\be
\label{112}
\Lambda(\vp,t) \simeq  -\gm_1 -\gm_3 -
\frac{2\gm_2\gm_3}{\gm_1 g\zeta} \qquad (g\zeta\gg 1) \; .
\ee
The stationary pumping parameter $\zeta$ is in the interval
$-1\leq\zeta\leq 1$, depending on the level of pumping. When there is
no stationary pumping, $\zeta=-1$. One says that the pumping is weak, if
$-1<\zeta<0$, and it is strong, if $0<\zeta<1$. Keeping in mind that the
coupling parameter $g$ is positive, we see that there exist two different
cases, when the stationary pumping is weak or absent, $\zeta<0$, and when
it is strong, $\zeta>0$. According to the principle of minimal expansion
(110), the minimum of the expansion rate corresponds to the maximum of
$g(\vp)$ if $\zeta<0$, and to the minimum of $g(\vp)$, if $\zeta>0$. The
extrema of $g(\vp)$, as follows from Eq. (105), are given by the equation
\be
\label{113}
{\rm Si}(2\vp) = \frac{\pi}{2} \; .
\ee

In the standard situation of absent or weak pumping, $\zeta<0$, we
have to look for the absolute maximum of $g(\vp)$. Then Eq. (113) gives
$\vp=0.96$. From relation (104), we have $R_n=0.55\sqrt{\lbd L}$, and
from Eq. (49), we find the {\it typical filament radius}
\be
\label{114}
r_f = 0.3\sqrt{\lbd L} \; .
\ee
The number of filaments can be estimated as $N_f\approx R^2/R_n^2$, which
yields
\be
\label{115}
N_f \approx 3.3 F \; .
\ee
The linear dependence of the filament number on the Fresnel number is
characteristic of the turbulent photon filamentation.

Note that under strong stationary pumping $(\zeta>0)$, when we need
to look for the minimum of $g(\vp)$, we would have $\vp=2.45$, hence,
$R_n=0.88\sqrt{\lbd L}$ and $r_f=0.5\sqrt{\lbd L}$.

The formula (114) for the typical filaments can be compared with the radii
observed in experiments. Thus, in different vapour lasers [25--29], one has
$r_f\approx 0.01$ cm. For the CO$_2$ laser and dye lasers, it was found
[36--42] that $r_f\approx 0.1$ cm and $r_f\approx 0.01$ cm, respectively.
All these data are in good agreement with formula (114).

It is also worth mentioning that a similar kind of turbulent photon
filamentation could arise in another type of matter, called photon band-gap
materials. These materials possess a prohibited band gap, where light cannot
propagate. The spontaneous radiation of atoms, with a frequency inside the
prohibited band gap, is strongly suppressed [64]. However, if the density
of doped atoms is sufficiently high, coherent interactions may develop (see
review [5]). Then atoms can start radiating even inside the prohibited band
gap. An unrealistic model, with the radiation length $\lbd\gg L$ much larger
than the system size was considered [65], where atoms with the resonance
frequency at the band edge could produce collective spontaneous emission.
Collective phenomena in the true atomic radiation of wavelength $\lbd\ll L$,
with the atomic frequency inside the prohibited band gap were also considered
and the effect of {\it collective liberation of light} was predicted [66--70].
Coherent radiation, accompanying this effect, should be realized by means of
a bunch of turbulent filaments.

\vskip 5mm

In conclusion, the theory of turbulent photon filamentation in large-aperture
lasers has been presented. This theory makes it possible to describe the
spatial filamentary stricture as well as its dynamics. The typical filament
radius, predicted by the theory, is in good agreement with experiments for
different lasers.

\vskip 5mm

{\bf Acknowledgement}

\vskip 2mm

I appreciate the Mercator Professorship from the German Research Foundation.

\end{document}